\documentclass[a4paper,conference]{IEEEtran}
\IEEEoverridecommandlockouts
\usepackage{cite}
\usepackage{amsmath,amssymb,amsfonts}
\usepackage{algorithmic}
\usepackage{graphicx}
\usepackage{textcomp}
\usepackage{xcolor}
\usepackage{booktabs}
\usepackage{amsthm}
\newtheorem{thm}{Theorem}
\newtheorem{corollary}{Corollary}
\newtheorem{lemma}{Lemma}
\newenvironment{prf}{{\indent \indent \it Proof:}}{\hfill $\blacksquare$\par}

\allowdisplaybreaks[3]
\newcounter{MYtempeqncnt}
\def\BibTeX{{\rm B\kern-.05em{\sc i\kern-.025em b}\kern-.08em
    T\kern-.1667em\lower.7ex\hbox{E}\kern-.125emX}}
\begin{document}

\title{Age of Information: A Two-Sensor Status
	Update System Monitoring The Same Process\\}

\author{
	\IEEEauthorblockN{Tianqing~Yang\IEEEauthorrefmark{1}, Zhengchuan~Chen\IEEEauthorrefmark{1}, Howard H. Yang\IEEEauthorrefmark{2}, Min Wang\IEEEauthorrefmark{3}, Yunjian~Jia\IEEEauthorrefmark{1}  and Tony Q. S. Quek\IEEEauthorrefmark{4}}
	
	\IEEEauthorblockA{\IEEEauthorrefmark{1}School of  Microelectronics and Communication Engineering, Chongqing University, Chongqing, China}
	\IEEEauthorblockA{\IEEEauthorrefmark{2}Zhejiang University/University of Illinois at Urbana-Champaign Institute, Zhejiang University, Zhejiang, China}
	\IEEEauthorblockA{\IEEEauthorrefmark{3}School of Optoelectronics Engineering, Chongqing University of Posts and Telecommunications, Chongqing, China}
	\IEEEauthorblockA{\IEEEauthorrefmark{4}Information System Technology and Design Pillar, Singapore University of Technology and Design, Singapore}
	Email: \IEEEauthorrefmark{1}\{ytq,~czc,~yunjian\}@cqu.edu.cn,~\IEEEauthorrefmark{2}haoyang@intl.zju.edu.cn,~\IEEEauthorrefmark{3}wangm@cqupt.edu.cn,~\IEEEauthorrefmark{4}tonyquek@sutd.edu.sg
}

\maketitle

\begin{abstract}
This work studies the average age of information (AoI) of a monitoring system in which two sensors are sensing the same physical process and update status to a common monitor using their dedicated channels.
Generally, using redundant devices to update the status of a process can improve the information timeliness at the monitor, but the disordered arrivals of updates also make the AoI analysis challenging.
To that end, we model the system as two M/M/1/1 parallel status updating queues. By leveraging tools from stochastic hybrid system (SHS), we provide a general approach to analyze the average AoI, whereas a closed-form expression can be obtained when the status arrival rates and/or the service rates are the same for the two sensors.
Numerical results validate the correctness of the theoretical analysis.
\end{abstract}

\begin{IEEEkeywords}
Age of information, stochastic hybrid systems, Internet of Things.
\end{IEEEkeywords}

\section{Introduction}
With the widespread deployment and adoption of Internet of things (IoT), it is critical to maintain the freshness of updated information for time sensitive applications, such as auto-driving vehicles and haptic communications.
To this end, age of information (AoI) has recently been proposed as a valid indicator of information freshness \cite{kaul2012real}.
In general, the AoI of the destination node at time $t$ is defined as $\Delta(t)=t-u(t)$, where $u(t)$ is the generation time of the received update \cite{kaul2012status}.
Different from traditional metrics such as delay or throughput, AoI is measured from the perspective of destination node \cite{bacinoglu2015age}.
Average AoI, defined as the time average of age, is one of the most commonly used AoI indicator.

In order to analyze and optimize AoI in wireless communication systems, a commonly adopted approach is to model the dynamics of AoI over a communication channel as a queuing system. Early works on AoI analysis focused on single-source state update systems \cite{kaul2012real, kaul2012status, costa2016age, najm2017status}.
Specifically, the authors explored M/M/1 queuing systems with first-come-first-served (FCFS) and last-come-first-served (LCFS) queuing disciplines in \cite{kaul2012real} and \cite{kaul2012status}, respectively, and derived general expressions for the corresponding average AoI.
\footnote{We follow the Kendall notaion system.}
The study in \cite{costa2016age} derived the general expression for the average AoI of the M/M/1/1 queuing system.
The authors in \cite{najm2017status} derived a general formula for the average AoI of M/G/1/1 non-preemptive systems.

There is a recent line of research that extends the AoI analysis to multi-source and multi-server systems \cite{yates2018age, akar2021discrete, yates2018status, zhou2020age,kalor2019minimizing}.
For example, the authors in \cite{yates2018age} used stochastic hybrid systems (SHS) to study AoI in multi-source state update systems for different queuing disciplines.
In \cite{akar2021discrete}, the authors proposed a discrete-time queuing model to derive the exact distribution of AoI sequences in a multi-source state update system with Bernoulli information packet arrival and discrete phase-type service times.
The study in \cite{yates2018status} analyzed the average AoI of parallel networks consisting of multiple memoryless preemptive LCFS queues, and derived the general expression for the average AoI of two server system with different service rates.
The authors in \cite{zhou2020age} investigated the optimal device scheduling process for a real-time IoT monitoring system with two devices and jointly minimized the average AoI of the destination and the energy cost of the device.
In \cite{kalor2019minimizing}, the authors investigated a data collection system composed of two related sources and improved the timeliness of information collection by optimizing the updating strategy for related sources.

In real-time IoT monitoring systems, the data arrival rate and the service rate of the sensor are two major factors affecting the freshness of received information at the monitor.
Due to the limited sensing and transmission capabilities of devices, using solely one sensor to update status risks the potential setback of encountering  long update intervals which deteriorates timeliness.
A rule of thumb in addressing this issue is to add redundant devices to monitor the same physical process can likely supply update more frequently and avoids the occurrence of long update intervals, hence, improving the data freshness.
Inspired by this, the present paper investigates the average AoI performance of a two-sensor parallel status updating system where a redundant sensor is employed to sense the same physical process and the two sensors update status using orthogonal channels.
In particular, considering storage limit of sensors and aiming at providing fresh update, the buffer at each sensor is restricted to one update.
Different from the study in \cite{yates2018status}, the system we consider consists of two M/M/1/1 non-preemptive parallel status updating queues.
It is noteworthy that, the parallel updating of sensors disrupts the causality of packet arrival at the monitor, which poses a significant challenge to the analysis of AoI.
In this work, we leverage the SHS framework to derive a general average AoI result for cases with different arrival rates and service rates.
Based on the general results, we further derive closed-form expressions for the average AoI with the same status arrival rates and/or service rates for the two-sensor parallel status updating system.

The remaining paper is organized as follows. 
The system model is described in Section II. 
In Section III, we detail the analysis of the average AoI of the system.
In Section IV, numerical results are provided to validate the correctness  of the theoretical analysis.
Finally, Section V concludes this paper.

\section{System Model}
\begin{figure}[htbp]
	\centering
	\includegraphics[width=2.7in]{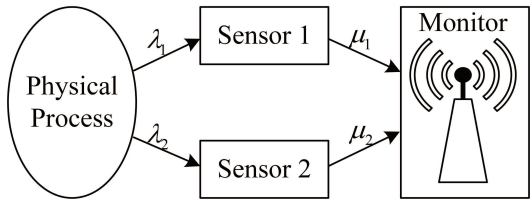}
	\caption{ The two-sensor status update system network model}
	\label{model}
\end{figure}

Consider a remote monitoring system consisting of two sensors and a monitor, as shown in Fig. \ref{model}.
Both sensors keep sampling the same physical process.
Specifically, the updates from sensor $i\in I:=\{1,2\}$, are generated according to a Poisson process of arrival rate $\lambda_{i}$, and subsequently sent to the monitor, in which updates are processed with service rate $\mu_{i}$ following a general Poisson distribution.
Noticing that the queue length might be limited for IoT based real-time systems, we consider each sensor operates under a unit size buffer with blocking discipline.
In other words, an update will only get served if it arrives when the system is idle, otherwise it is discarded directly.
	
In the absence of any updates, the instantaneous age of the monitor increases linearly with time.
Moreover, due to the randomness of service times and reach times, some updates may be out of date when they arrive at the monitor. 
Therefore, when a new update arrives at the monitor, the monitor compares the instantaneous age with the age of the newly arrived update.
And only when the age of the newly arrived update is less than the instantaneous age of the monitor, the monitor will adopt the update.
Consequently, we model the system as two M/M/1/1 parallel status updating queues.		
\section{Age Average of the Two-Sensor Status Update System}
Since with parallel updating, arrival of updates at the monitor would be disorder, it is not straightforward to describe the AoI process with commonly used graph-based geometric analysis.
In this section, we leverage the SHS framework introduced in \cite{yates2018age} to derive the average AoI of the considered system.
We first briefly introduce SHS and then apply the SHS to analyze the average AoI of the two-sensor status update system.

\subsection{Stochastic Hybrid Systems for AoI}
In general, an SHS is defined as a hybrid system consisting of a continuous and a discrete part where the two parts are stochastic components.
Specifically, the discrete state at time $t$ is denoted by $q(t) \in \mathcal{Q}$, which can be modeled by a Markov chain.
The continuous part is denoted by $\boldsymbol{x}(t)=\left[x_{0}(t), x_{1}(t), \cdots, x_{n}(t)\right]$ where $x_{i}(t), \ i=1,2, \cdots, n$ represents continuous processes. 
To characterize the AoI of the status update, one generally considers that the discrete part represents the buffer state, while the continuous process consists of age-related processes.

In the SHS, the Markov chain $q(t)$ is further clarified with transition set $\mathcal{L}$.
Specifically, when a status update is stored the buffer or finishes transmission, a transition between states would occur.
For a transition $l \in \mathcal{L}$, the current state and the next state before and after the transition are denoted by $q_{l}$ and $q_{l}^{\prime}$, respectively.
The transition rate $\lambda^{(l)}$ describes the continuous-time Markov chain for $q(t)$.
A transition $l$ would cause a jump from discrete state $q_{l}$ and reset the continuous state from $\boldsymbol{x}$ to $\boldsymbol{x}^{\prime}=\boldsymbol{x} \boldsymbol{A}_{l}$  where $\boldsymbol{A}_{l} \in \mathbb{R}^{n+1} \times \mathbb{R}^{n+1}$ is the reset map of transition $l$. 

In a specific state $q \in \mathcal{Q}$, the continuous state evolves according to $\dot{\boldsymbol{x}}(t)=\mathbf{b}_{q}$, where the $i$-th element of $b_q$, $b_{q}^{(i)}$ represents the slope of $x_i(t)$.
Particularly, the elements of $\mathbf{b}_{q}$ is binary when we use SHS tools for AoI.
The age process $x_{i}(t)$ increases at a unit rate when $b_{q}^{(i)}$ is equal to 1.
When $b_{q}^{(i)}$ is equal to 0, the age process $x_{i}(t)$ keeps the same value.

The probability of a Markov chain being in state $q$ is denoted as $\pi_{q}(t)$.
The correlation vector between $q(t)$ and $\boldsymbol{x}(t)$ is denoted as $\boldsymbol{v}_{q}(t)$.
We formally define $\pi_{q}(t)$ and $\boldsymbol{v}_{q}(t)$ by the following equations,
\begin{equation}
	\pi_{q}(t)=\mathbb{E}\left[\delta_{q, q(t)}\right]=P(q(t)=q),\label{define of pi}
\end{equation}
\begin{equation}
	\boldsymbol{v}_{q}(t)=\left[v_{q 0}(t) \quad v_{q 1}(t) \ldots v_{q n}(t)\right]=\mathbb{E}\left[\boldsymbol{x} \delta_{q, q(t)}\right],\label{define of v}
\end{equation}
where $\delta_{q, q(t)}$ denotes the Kronecker delta function, i.e., $\delta_{q, q(t)}=1$ if $q_{t}=q$, otherwise  $\delta_{q, q(t)}=0$.
With the definition of $\boldsymbol{v}_{q}(t)$, it has that $\mathbb{E}[\boldsymbol{x}(t)]=\sum_{q \in \mathcal{Q}} \lim _{t \rightarrow \infty} \boldsymbol{v}_{q}(t)$.

In fact, following the ergodicity, the stationary distribution of the states exists uniquely and the state probability $\pi_{q}(t)$ would converge to its stationary counterpart $\bar{\pi}_{q}$ for $q \in \mathcal{Q}$.
Similarly, the correlation vector $\boldsymbol{v}_{q}$ would also converge to a nonnegative vector $\overline{\boldsymbol{v}}_{q}$ for $q \in \mathcal{Q}$. 
Thus, one has that
\begin{equation}
	\mathbb{E}[\boldsymbol{x}(t)]=\sum_{q \in \mathcal{Q}} \lim _{t \rightarrow \infty} \boldsymbol{v}_{q}(t)=\sum_{q \in \mathcal{Q}} \overline{\boldsymbol{v}}_{q}.\label{ext}
\end{equation}
That is, to know the average of an age process, it is equivalent to get the corresponding average correlation limit for all states.

Define the outgoing transition set and incoming transition set for a state $q$ as
\begin{equation}
	\mathcal{L}_{q}^{\prime}=\left\{l \in \mathcal{L}: q_{l}^{\prime}=q\right\}, \quad \mathcal{L}_{q}=\left\{l \in \mathcal{L}: q_{l}=q\right\}.
\end{equation}
A general property has been known for the presented SHS model.
We provide it as the following lemma.
\begin{lemma} \cite[Appendix C]{yates2018age}
	\label{lemma1}
	In an SHS ($q(t),\boldsymbol{x}(t)$) for $q \in \mathcal{Q}$, the steady state probability $\bar{\pi}_{q}$ and vector function $\overline{\boldsymbol{v}}_{q}$ satisfy	the following equations:
\begin{align}
		\bar{\pi}_{q} \sum_{l \in \mathcal{L}_{q}} \lambda^{(l)}=&\sum_{l \in \mathcal{L}_{q}^{\prime}} \lambda^{(l)} \bar{\pi}_{q_{l}}, \quad q \in \mathcal{Q}\label{pi_1},\\
		\sum_{q \in \mathcal{Q}} \bar{\pi}_{q}=&1,\label{pi_2}\\
	    \overline{\boldsymbol{v}}_{q}\left(\sum_{l \in \mathcal{L}_{q}} \lambda^{(l)}\right)=&\boldsymbol{b}_{q} \bar{\pi}_{q}+\sum_{l \in \mathcal{L}_{q}^{\prime}} \lambda^{(l)} \overline{\boldsymbol{v}}_{q l} \boldsymbol{A}_{l}, \quad q \in \mathcal{Q}.\label{vq}
\end{align}
\end{lemma}

According to Lemma \ref{lemma1}, one can get the stationary distribution $\bar{\pi}:=\left(\bar{\pi}_{0}, \bar{\pi}_{1}, \cdots, \bar{\pi}_{n}\right)$ from (\ref{pi_1})-(\ref{pi_2}). Then, one can derive all the average correlation limit $\left\{\overline{\boldsymbol{v}}_{q},  q \in \mathcal{Q}\right\}$ from $\bar{\pi}$ and (\ref{vq}). Accordingly, the average of age-related processes can be obtained from $\left\{\overline{\boldsymbol{v}}_{q}, q \in \mathcal{Q}\right\}$ and (\ref{ext}).

\subsection{ Average AoI of the Two-Sensor Status Update System}

Based on the SHS framework, one can establish appropriate state set $\mathcal{Q}$ and age-related process $\boldsymbol{x}(t)$ in the two-sensor parallel status updating system to derive the average AoI.
In consequence, one can derive analytical expressions for the average AoI as follows.
\begin{thm}\label{thm1}
	In the two-sensor status update system, the average AoI is equal to $\overline{\Delta}=\sum_{q \in Q} \bar{v}_{q 0}$ with $\overline{\boldsymbol{v}}=\left[\overline{\boldsymbol{v}}_{0} , \overline{\boldsymbol{v}}_{1},\cdots,\overline{\boldsymbol{v}}_{8} \right]$ being the solution of the following equations:
	\begin{subequations}
		\begin{align}
			(\lambda_{1}+\lambda_{2})\overline{\boldsymbol{v}}_{0}=&\bar{\pi}_{0}[1,1,1]+\mu_{1}\left[\bar{v}_{11}, 0, 0\right]+\mu_{1}\left[\bar{v}_{20}, 0,0\right]+\notag\\
			&\mu_{2}\left[\bar{v}_{42}, 0,0\right]+\mu_{2}\left[\bar{v}_{50}, 0,0\right],\label{v_a}\\
			(\lambda_{2}+\mu_{1})\overline{\boldsymbol{v}}_{1}    =&\bar{\pi}_{1}[1,1,0]+\lambda_{1}\left[\bar{v}_{00}, 0, 0\right]+\mu_{2}\left[\bar{v}_{62},\bar{v}_{61}, 0\right]+\notag\\
			&\mu_{2}\left[\bar{v}_{80},\bar{v}_{81}, 0\right],\\
			(\lambda_{2}+\mu_{1})\overline{\boldsymbol{v}}_{2}    =&\bar{\pi}_{2}[1,1,0]+\mu_{2}\left[\bar{v}_{32}, \bar{v}_{31}, 0\right]+\mu_{2}\left[\bar{v}_{72}, \bar{v}_{71}, 0\right],\\
			(\mu_{1}+\mu_{2})\overline{\boldsymbol{v}}_{3}        =&\bar{\pi}_{3}[1,1,1]+\lambda_{2}\left[\bar{v}_{10}, \bar{v}_{11}, 0\right],\\
			(\lambda_{1}+\mu_{2})\overline{\boldsymbol{v}}_{4}    =&\bar{\pi}_{4}[1,0,1]+\lambda_{2}\left[\bar{v}_{00}, 0, 0\right]+\mu_{1}\left[\bar{v}_{31}, 0, \bar{v}_{32}\right]+\notag\\
			&\mu_{1}\left[\bar{v}_{70}, 0, \bar{v}_{72}\right],\\
			(\lambda_{1}+\mu_{2})\overline{\boldsymbol{v}}_{5}    =&\bar{\pi}_{5}[1,0,1]+\mu_{1}\left[\bar{v}_{61}, 0, \bar{v}_{62}\right]+\mu_{1}\left[\bar{v}_{81}, 0, \bar{v}_{68}\right],\\
			(\mu_{1}+\mu_{2})\overline{\boldsymbol{v}}_{6}        =&\bar{\pi}_{6}[1,1,1]+\lambda_{1}\left[\bar{v}_{40}, 0, \bar{v}_{42}\right],\\
			(\mu_{1}+\mu_{2})\overline{\boldsymbol{v}}_{7}        =&\bar{\pi}_{7}[1,1,1]+\lambda_{2}\left[\bar{v}_{20}, \bar{v}_{21}, 0\right],\\
			(\mu_{1}+\mu_{2})\overline{\boldsymbol{v}}_{8}        =&\bar{\pi}_{8}[1,1,1]+\lambda_{1}\left[\bar{v}_{50}, 0, \bar{v}_{52}\right],\label{v_i}
		\end{align}
		\label{v}
	\end{subequations}
where 
\begin{subequations}
	\begin{align}
		&\bar{\pi}_{0}=\frac{\mu_{1}\mu_{2}(\mu_{1}+\mu_{2})^{2}}{G(\lambda_{1},\lambda_{2},\mu_{1},\mu_{2})}\label{solution_pi_0},\\
		&\bar{\pi}_{1}=\frac{\lambda_{1}\mu_{1}\mu_{2}(\mu_{1}+\mu_{2})(\lambda_{2}+\mu_{1}+\mu_{2})}{(\lambda_{2}+\mu_{1})G(\lambda_{1},\lambda_{2},\mu_{1},\mu_{2})},\\
		&\bar{\pi}_{2}=\frac{\lambda_{1}\lambda_{2}\mu_{2}^{2}(\mu_{1}+\mu_{2})}{(\lambda_{2}+\mu_{1})G(\lambda_{1},\lambda_{2},\mu_{1},\mu_{2})},\\
		&\bar{\pi}_{3}=\frac{\lambda_{1}\lambda_{2}\mu_{1}\mu_{2}(\lambda_{2}+\mu_{1}+\mu_{2})}{(\lambda_{2}+\mu_{1})G(\lambda_{1},\lambda_{2},\mu_{1},\mu_{2})},\\
		&\bar{\pi}_{4}=\frac{\lambda_{2}\mu_{1}\mu_{2}(\mu_{1}+\mu_{2})(\lambda_{1}+\mu_{1}+\mu_{2})}{(\lambda_{1}+\mu_{2})G(\lambda_{1},\lambda_{2},\mu_{1},\mu_{2})},\\
		&\bar{\pi}_{5}=\frac{\lambda_{1}\lambda_{2}\mu_{1}^{2}(\mu_{1}+\mu_{2})}{(\lambda_{1}+\mu_{2})G(\lambda_{1},\lambda_{2},\mu_{1},\mu_{2})},\\
		&\bar{\pi}_{6}=\frac{\lambda_{1}\lambda_{2}\mu_{1}\mu_{2}(\lambda_{1}+\mu_{1}+\mu_{2})}{(\lambda_{1}+\mu_{2})G(\lambda_{1},\lambda_{2},\mu_{1},\mu_{2})},\\
		&\bar{\pi}_{7}=\frac{\lambda_{1}\lambda_{2}^{2}\mu_{2}^{2}}{(\lambda_{2}+\mu_{1})G(\lambda_{1},\lambda_{2},\mu_{1},\mu_{2})},\\
		&\bar{\pi}_{8}=\frac{\lambda_{1}^{2}\lambda_{2}\mu_{1}^{2}}{(\lambda_{1}+\mu_{2})G(\lambda_{1},\lambda_{2},\mu_{1},\mu_{2})}\label{solution_pi_8},
	\end{align}
\end{subequations}
 and
\begin{equation}
G(\lambda_{1},\lambda_{2},\mu_{1},\mu_{2})=(\lambda_{1}+\mu_{1})(\lambda_{2}+\mu_{2})(\mu_{1}+\mu_{2})^{2}. 
\end{equation}
\end{thm}

\begin{prf}
The first step of our proof consists of defining the discrete states set  $\mathcal{Q}$.
We divide the $q(t)$ into nine states as $\mathcal{Q}=\{0,1,2,3,4,5,6,7,8\}$ where 
\begin{itemize}
	\item $q(t)=0$ indicates that no update is under transmission.
	\item $q(t)=1$ represents that an update generated from sensor 1 is under transmission.
	No updates have been generated from sensor 2 and then sent to the monitor since the update generated from sensor 1 was generated.
	\item $q(t)=2$ represents that an update generated from sensor 1 is under transmission.
	Since sensor 1 generated the update, sensor 2 generated the update and was sent to the monitor.
	\item $q(t)=3$ represents that an update generated from sensor 1 and an update generated from sensor 2 are being transmitted, while the update from sensor 2 is generated later than the update from sensor 1, no updates have been generated from sensor 2 and then sent to the monitor since the update generated from sensor 1 was generated.
	\item $q(t)=4$ represents that an update generated from sensor 2 is under transmission, while no updates have been generated from sensor 1 and then sent to the monitor since the update generated from sensor 2 was generated.
	\item $q(t)=5$ represents that an update generated from sensor 2 is under transmission, while the sensor 1 has generated the update, which is then sent to the monitor since the update generated from sensor 2 was generated.
	\item $q(t)=6$ represents that an update generated from sensor 1 and an update generated from sensor 2 are being transmitted, while the update from sensor 2 is generated earlier than the update from sensor 1, no updates have been generated from sensor 1 and then sent to the monitor since the update generated from sensor 2 was generated.
	\item $q(t)=7$ represents that represents that an update generated from sensor 1 and an update generated from sensor 2 are being transmitted, while the update from sensor 2 is generated later than the update from sensor 1. From the time the update generated by sensor 1 is produced since until now, sensor 2 generated the update and then sent it to the monitor.
	\item $q(t)=8$ represents that an update generated from sensor 1 and an update generated from sensor 2 are being transmitted, while the update from sensor 2 is generated earlier than the update from sensor 1. 
	From the moment the update is generated by sensor 2, sensor 1 generated the update and then sent it to the monitor.
\end{itemize}

The continuous-time state process is $\boldsymbol{x}(t)=\left[x_{0}(t) , x_{1}(t) , x_{2}(t)\right]$ where $x_{0}(t)$ is the age of the monitor at time $t$ and $x_{i}(t)$ is the age of the packet in the sensor $i$ at time $t$.
The corresponding Markov chain with state space $\mathcal{Q}$ for the SHS is shown in Fig. \ref{status_figure}. 
\begin{figure}[htbp]
	\centering
	\includegraphics[height=5cm,width=5cm]{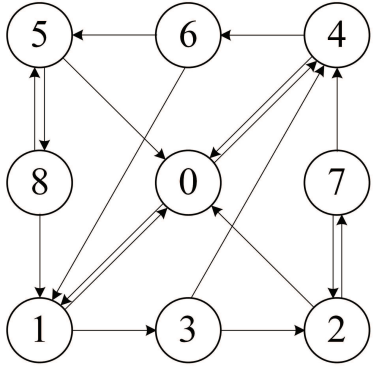}
	\caption{The SHS Markov chain for the two-sensor status update system}
	\label{status_figure}
\end{figure} 
The transition rate, the value of continuous process after the transition, and the value of correlation function after the transition maps are summarized as Table I.
\begin{table}[htbp]  
	\caption{Table of transitions for the Markov chain in Figure 2.}   
	\centering
	\renewcommand{\arraystretch}{1.5} 
	\begin{tabular*}{6.5cm}{ccccc}
		\bottomrule
		$l$   & $q_{l} \rightarrow q_{l}^{\prime}$ & $\lambda^{(l)}$ & $\boldsymbol{x} \boldsymbol{A}_{l}$ & $\boldsymbol{v}_{q_{l}} \boldsymbol{A}_{l}$\\[2.5pt]  
		\hline  
		$1$   & $0 \rightarrow 1$ & $\lambda_{1}$ & $ \left[x_{0},0,0\right]$                                     & $\left[v_{00}, 0, 0\right]$      \\[7.5pt] 
		$2$   & $1 \rightarrow 0$ & $\mu_{1}$     & $ \left[x_{1},0,0\right]$                                     & $\left[v_{11}, 0, 0\right]$      \\[7.5pt] 
		$3$   & $1 \rightarrow 3$ & $\lambda_{2}$ & $ \left[x_{0},x_{1},0\right]$                                   & $\left[v_{10}, v_{11}, 0\right]$ \\[7.5pt] 
		$4$   & $3 \rightarrow 2$ & $\mu_{2}$     & $ \left[x_{2},x_{1}, 0\right]$                                   & $\left[v_{32}, v_{31}, 0\right]$ \\[7.5pt]
		$5$   & $2 \rightarrow 0$ & $\mu_{1}$     & $ \left[x_{0},0, 0\right]$                                       & $\left[v_{20}, 0, 0\right]$      \\[7.5pt]  
		$6$   & $2 \rightarrow 7$ & $\lambda_{2}$ & $ \left[x_{0},x_{1}, 0\right]$                                   & $\left[v_{20},v_{21}, 0\right]$  \\[7.5pt] 
		$7$   & $7 \rightarrow 2$ & $\mu_{2}$     & $ \left[x_{2},x_{1}, 0\right]$                                   & $\left[v_{72},v_{71}, 0\right]$  \\[7.5pt] 
		$8$   & $3 \rightarrow 4$ & $\mu_{1}$     & $ \left[x_{1},0, x_{2}\right]$                                   & $\left[v_{31},0, v_{32}\right]$  \\[7.5pt] 
		$9$   & $7 \rightarrow 4$ & $\mu_{1}$     & $ \left[x_{0},0, x_{2}\right]$                                   & $\left[v_{70},0, v_{72}\right]$  \\[7.5pt] 
		$10$  & $0 \rightarrow 4$ & $\lambda_{2}$ & $ \left[x_{0},0, 0\right]$                                       & $\left[v_{00},0, 0\right]$       \\[7.5pt] 
		$11$  & $4 \rightarrow 0$ & $\mu_{2}$     & $ \left[x_{2},0, 0\right]$                                       & $\left[v_{42},0, 0\right]$       \\[7.5pt] 
		$12$  & $4 \rightarrow 6$ & $\lambda_{1}$ & $ \left[x_{0},0, x_{2}\right]$                                   & $\left[v_{40},0, v_{42}\right]$  \\[7.5pt]
		$13$  & $6 \rightarrow 5$ & $\mu_{1}$     & $ \left[x_{1},0, x_{2}\right]$                                   & $\left[v_{61},0, v_{62}\right]$  \\[7.5pt]
		$14$  & $5 \rightarrow 0$ & $\mu_{2}$     & $ \left[x_{0},0, 0\right]$                                       & $\left[v_{50},0, 0\right]$       \\[7.5pt]
		$15$  & $5 \rightarrow 8$ & $\lambda_{1}$ & $ \left[x_{0},0, x_{2}\right]$                                   & $\left[v_{50},0, v_{52}\right]$  \\[7.5pt]
		$16$  & $8 \rightarrow 5$ & $\mu_{1}$     & $ \left[x_{1},0, x_{2}\right]$                                   & $\left[v_{81},0, v_{82}\right]$  \\[7.5pt]
		$17$  & $6 \rightarrow 1$ & $\mu_{2}$     & $ \left[x_{2},x_{1}, 0\right]$                                   & $\left[v_{62},v_{61}, 0\right]$  \\[7.5pt]
		$18$  & $8 \rightarrow 1$ & $\mu_{2}$     & $ \left[x_{0},x_{1}, 0\right]$                                   & $\left[v_{80},v_{81}, 0\right]$  \\[7.5pt]
		\bottomrule		
	\end{tabular*} 
\end{table} 

In the following, we describe the transitions in Table I:
\begin{itemize}
		\item $l=1$: A new status update generated from sensor 1 arrives at the transmitter.
		Hence, the transition rate is $\lambda_{1}$.
		Along with this update arrival, the AoI of monitor remains the same, i.e., $x_{0}^{\prime}=x_{0}$, since the newly arrived status update has been delivered to the monitor.
		Moreover, $x_{1}^{\prime}=0$, since the newly arrived status update is just generated.
		Since $x_{2}$ is independent of $x_{1}$, $x_{2}^{\prime}=0$.
		Thus, we have
		\begin{equation}
			\boldsymbol{x}^{\prime}=\left[x_{0}, x_{1}, x_{2}\right] \boldsymbol{A}_{1}=\left[x_{0}, 0,0\right] .
		\end{equation}
     	Then, one can get $\boldsymbol{A}_{1}$ as 
     	\begin{equation}
     		\boldsymbol{A}_{1}=\begin{bmatrix}
     		1	& 0 & 0\\
     		0	& 0 & 0\\
     		0	& 0 &0
     		\end{bmatrix}.
     	\end{equation}
        Based on the reset matrix $\boldsymbol{A}_{1}$, one can further compute $\boldsymbol{v}_{0} \boldsymbol{A}_{1}$ as
        \begin{equation}
        	\overline{\boldsymbol{v}}_{0} \boldsymbol{A}_{1}=\left[v_{00}, v_{01}, v_{02}\right] \boldsymbol{A}_{1}=\left[v_{00}, 0,0\right].
        \end{equation}
        Since if one can write $\boldsymbol{x}^{\prime}$, then $\boldsymbol{A}_{l}$ and $\boldsymbol{v}_{q_{l}} \boldsymbol{A}_{l}$ are easy to calculate.
        For the rest of the transitions, we focus on the analysis of $\boldsymbol{x}^{\prime}$ and omit the analysis of $\boldsymbol{A}_{l}$ and $\boldsymbol{v}_{q_{l}} \boldsymbol{A}_{l}$.
		The situation is similar when $l=9$.
        
        \item $l=2$: An update is sent by sensor 1. 
        In this case, the transition rate is $\mu_{1}$. 
        $x_{0}^{\prime}=x_{1}$ because no updates have been generated from sensor 2 and sent to the monitor since sensor 1 generated an update which means the age of newly
        arrived update sent by sensor 1 is less than the instantaneous age at the monitor. 
        Besides, $x_{1}^{\prime}=0$ and $x_{2}^{\prime}=0$ since there would be no update from sourse 1 under transmission and $x_{2}$ is independent of $x_{1}$.
        The situation is similar when $l=10$.
        
        \item $l=3$: A new status update generated from sensor 2 arrives at the transmitter.
        The transition rate is $\lambda_{2}$.
        Note that in this case $x_{0}^{\prime}=x_{0}$, $x_{1}^{\prime}=x_{1}$ and $x_{2}^{\prime}=0$ since the arrived status update has not been delivered to the monitor and $x_{1}$ is independent of $x_{2}$.
        The situation is similar when $l=11$.
        
        \item $l=4$: An update of sensor 2 finishes transmission with service rate $\mu_{2}$.
        $x_{0}^{\prime}=x_{2}$ since the newly arrived update is generated later than the update generation from sensor 1.
        Besides, $x_{1}^{\prime}=x_{1}$ and $x_{2}^{\prime}=0$ since there would be no update from sourse 2 under transmission and $x_{1}$ is independent of $x_{2}$.
        The situation is similar when $l=12$.
        
        \item $l=5$: An update is sent by sensor 2.
        Hence, the transition rate is $\mu_{2}$.
        In this case, $x_{0}^{\prime}=x_{0}$ because after generating the update from sensor 1, sensor 2 generated the update and was sent to the monitor, which means the age of the newly arrived update sent by sensor 1 is larger than the instantaneous age at the monitor. 
        $x_{1}^{\prime}=0$ and $x_{2}^{\prime}=0$ since there would be no update under transmission.
        The situation is similar when $l=13$.
        
        \item $l=5$: An update is sent by sensor 1.
        Hence, the transition rate is $\mu_{1}$.
        In this case, $x_{0}^{\prime}=x_{0}$ because after generating the update from sensor 1, sensor 2 generated the update and was sent to the monitor, which means the age of the newly arrived update sent by sensor 1 is larger than the instantaneous age at the monitor. 
        $x_{1}^{\prime}=0$ and $x_{2}^{\prime}=0$ since there would be no update under transmission.
        The situation is similar when $l=14$.
        
        \item $l=6$: A new status update generated from sensor 2 arrives at the transmitter.
        the transition rate is $\lambda_{2}$.
        Along with this update arrival,  $x_{0}^{\prime}=x_{0}$, $x_{1}^{\prime}=x_{1}$ and $x_{2}^{\prime}=0$ since the arrived status update has not been delivered to the monitor and $x_{1}$ is independent of $x_{2}$.
        The situation is similar when $l=15$.
        
        \item $l=7$: An update of sensor 2 finishes transmission with service rate $\mu_{2}$.
        In this case, $x_{0}^{\prime}=x_{2}$ since the newly arrived update is generated later than the update generation from sensor 1.
        $x_{1}^{\prime}=x_{1}$ and $x_{2}^{\prime}=0$ since the update generated from sensor 1 is under transmission.
        The situation is similar when $l=16$.
        
        \item $l=8$: An update of sensor 1 finishes transmission with service rate $\mu_{1}$.
        $x_{0}^{\prime}=x_{1}$  because no updates have been generated from sensor 2 and sent to the monitor since sensor 1 generated an update which means the age of the newly arrived update sent by sensor 1 is less than the instantaneous age at the monitor. 
        Besides, $x_{1}^{\prime}=0$ and $x_{2}^{\prime}=x_{2}$ since there would be no update from sourse 1 under transmission and $x_{2}$ is independent of $x_{1}$.
        The situation is similar when $l=17$.
        
        \item $l=9$: An update is sent by sensor 1.
        Hence, the transition rate is $\mu_{1}$.
        Note that in this case, $x_{0}^{\prime}=x_{1}$ because no updates have been generated from sensor 2 and sent to the monitor since sensor 1 generated an update which indicates the age of the newly arrived update sent by sensor 1 is less than the instantaneous age at the monitor. 
        $x_{1}^{\prime}=0$ and $x_{2}^{\prime}=x_{2}$ since the update generated from sensor 2 is under transmission.
        The situation is similar when $l=18$.      
\end{itemize}

With the specified transition rates, the stationary probability $\boldsymbol{\bar{\pi}}$ can be obtained.
According to (\ref{pi_1}) and (\ref{pi_2}), $\boldsymbol{\bar{\pi}}$ satisfies that
\begin{subequations}
	\begin{align}
		&(\lambda_{1}+\lambda_{2})\bar{\pi}_{0}=\mu_{1}\bar{\pi}_{1}+\mu_{1}\bar{\pi}_{2}+\mu_{2}\bar{\pi}_{4}+\mu_{2}\bar{\pi}_{5},\label{pi_a}\\
		&(\lambda_{2}+\mu_{1})\bar{\pi}_{1}=\lambda_{1}\bar{\pi}_{0}+\mu_{2}\bar{\pi}_{6}+\mu_{2}\bar{\pi}_{8},\\
		&(\lambda_{2}+\mu_{1})\bar{\pi}_{2}=\mu_{2}\bar{\pi}_{3}+\mu_{2}\bar{\pi}_{7},\\
		&(\mu_{1}+\mu_{2})\bar{\pi}_{3}=\lambda_{2}\bar{\pi}_{1},\\
		&(\lambda_{1}+\mu_{2})\bar{\pi}_{4}=\lambda_{2}\bar{\pi}_{0}+\mu_{1}\bar{\pi}_{3}+\mu_{1}\bar{\pi}_{7},\\
		&(\lambda_{1}+\mu_{2})\bar{\pi}_{5}=\mu_{1}\bar{\pi}_{6}+\mu_{1}\bar{\pi}_{8},\\
		&(\mu_{1}+\mu_{2})\bar{\pi}_{6}=\lambda_{1}\bar{\pi}_{4},\\
		&(\mu_{1}+\mu_{2})\bar{\pi}_{7}=\lambda_{2}\bar{\pi}_{2},\\
		&(\mu_{1}+\mu_{2})\bar{\pi}_{8}=\lambda_{1}\bar{\pi}_{5}.\label{pi_i}
	\end{align}
\end{subequations}
Then, one can get $\boldsymbol{\bar{\pi}}$ from (\ref{pi_a})-(\ref{pi_i}) as (\ref{solution_pi_0}-\ref{solution_pi_8}).

By definition, $x_{0}$ always increases at a unit rate in each state, and $x_{i}$ grows linearly with rate 1 if there exists an update from sensor $i$ at the transmitter. 
Therefore, $\boldsymbol{b}_{q}$ can be expressed as
\begin{subequations}
	\begin{align}
		&\boldsymbol{b}_{q}=[1,0,0], \quad q \in\{0\},\\
		&\boldsymbol{b}_{q}=[1,1,0], \quad q \in\{1,2\},\\
		&\boldsymbol{b}_{q}=[1,0,1], \quad q \in\{4,5\},\\
		&\boldsymbol{b}_{q}=[1,1,1], \quad q \in\{3,6,7,8\}.
	\end{align}
\label{bq}
\end{subequations}
By applying (\ref{vq}), we can obtain $\overline{\boldsymbol{v}}=\left[\overline{\boldsymbol{v}}_{0} , \overline{\boldsymbol{v}}_{1},\cdots,\overline{\boldsymbol{v}}_{9} \right]$ by solving equation (\ref{v}).
Finally, we can get the average AoI of the two-sensor status update system by (\ref{ext}).
\end{prf}

In the special case where the channels between each sensor and the monitor has the same characteristics, namely they follow the same statistics of channel gain, the service rate of both sensors would be the same.
This case is very common in many pracitical IoT applications. In this case, $\mu_1=\mu_2$ and one can obtain the explicit expression of the average AoI as summarized in the following corollary.
\begin{corollary}
	When $\mu_{1}=\mu_{2}=\mu$, the average AoI of the two-sensor status update system is given by (\ref{same_mu}), at the top of next page.
\end{corollary}
\begin{figure*}[!t]
	\normalsize
	\setcounter{MYtempeqncnt}{\value{equation}}
	\begin{equation}
		\begin{aligned}
			\label{same_mu}
			&\overline{\Delta}=\frac{\lambda _1^4 \left(17 \lambda _2 \mu ^2+15 \lambda _2^2 \mu \!+\!5 \lambda _2^3+8 \mu ^3\right)\!+\!4 \mu ^3 \left(\lambda _2+\mu \right){}^2 \left(2 \lambda _2 \mu +2 \lambda _2^2+\mu ^2\right)\!+\!\lambda _1^3 \left(59 \lambda _2 \mu ^3+62 \lambda _2^2 \mu ^2+30 \lambda _2^3 \mu +5 \lambda _2^4\right)}{4 \left(\lambda _1+\lambda _2\right) \mu  \left(\lambda _1+\mu \right){}^3 \left(\lambda _2+\mu \right){}^3}\\
			&+\frac{24 \lambda _1^3\mu ^4+\lambda _1^2 \mu  \left(82 \lambda _2 \mu ^3+102 \lambda _2^2 \mu ^2+62 \lambda _2^3 \mu +15 \lambda _2^4+28 \mu ^4\right)+\lambda _1 \mu ^2 \left(56 \lambda _2 \mu ^3+82 \lambda _2^2 \mu ^2+59 \lambda _2^3 \mu +17 \lambda _2^4+16 \mu ^4\right)}{4 \left(\lambda _1+\lambda _2\right) \mu  \left(\lambda _1+\mu \right){}^3 \left(\lambda _2+\mu \right){}^3}
		\end{aligned}	
	\end{equation}
	\setcounter{equation}{\value{MYtempeqncnt}}
	\hrulefill
	\vspace*{4pt}
\end{figure*}
In a more homogeneous case where the sensors come from the same model, the arrival rates of the two sensors would also the same. Accordingly, based on Corolloary 1, we have the following result. 
\begin{corollary}
	 When $\lambda_{1}=\lambda_{2}=\lambda$ , $\mu_{1}=\mu_{2}=\mu$, the average AoI of the two-sensor status update system is 
	\begin{equation}
		\setcounter{equation}{17}
			\overline{\Delta}=\frac{5 \lambda ^5+20 \lambda ^4 \mu +34 \lambda ^3 \mu ^2+30 \lambda ^2 \mu ^3+12 \lambda  \mu ^4+2 \mu ^5}{4 \lambda  \mu  (\lambda +\mu )^4}.\label{allsame}
	\end{equation}
\end{corollary}
To improve the freshness of the information, it is usually achieved by eliminating the waiting time in the queue.
In this case, when $\lambda_{i} \rightarrow \infty$, $\overline{\Delta}=\frac{5}{4 \mu}$.
This result is the average AoI of the two-sensor status update system with a Zero-Wait (ZW) strategy.

\section{Numerical Results}
\begin{figure}[htbp]
	\centering
	\includegraphics[width=3in]{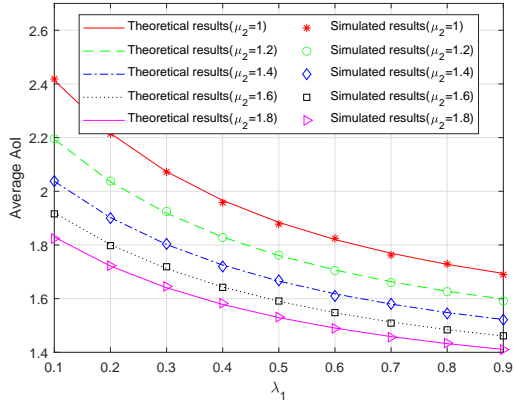}
	\caption{Average AoI of the two-sensor status update system, where $\lambda_{2}=0.8$ and $\mu_{1}=1$.}
	\label{yanzheng}
\end{figure}

In this section, some numerical results based on the theoretical analysis are provided to evaluate the performance of the two-sensor status update system.
First, we provide simulation results to verify the reliability of the theoretical analysis of the system under consideration.
Fig. \ref{yanzheng} shows the theoretical and simulated average AoI of the considered system when $\mu_{2}$ grows from 1 to 1.8 and $\lambda_{1}$ grows from 0.1 to 0.9 for $\mu_{1}=1$ and $\lambda_{2}=0.8$.
We have also simulated the system and computed the age over $2\times10^{5}$ time units and averaged over multiple trials, and the result is very close to the numerically-evaluated theoretical age. 
It is noted that the average AoI decreases rapidly as $\mu_{2}$ and $\lambda_{1}$ increase since more frequent transmissions leads to more frequent updates in this system model.

\begin{figure}[htbp]
	\centering
	\includegraphics[width=3in]{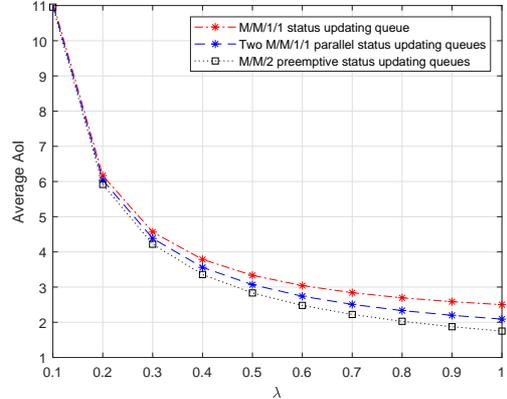}
	\caption{ Average AoI vs. arrival rate for M/M/1/1 status updating queue, two M/M/1/1 parallel status updating queues, and M/M/2 preemptive status updating queues, $\mu=1$.}
	\label{duibi2}
\end{figure}

In Fig. \ref{duibi2}, we have plotted the average age for two M/M/1/1 parallel status updating queues ($\mu_{1}=\mu_{2}=\mu$, $\lambda_{1}=\lambda_{2}=\lambda$/2), M/M/1/1 status updating queue, and M/M/2 preemptive status updating queues (see \cite[Sec. III]{yates2018status}) models as a function of the arrival rate $\lambda$ for the case $\mu=1$.
It can be seen that the AoI performance of our proposed model is better than the M/M/1/1 status updating queue model regardless of the arrival rate.
Since we add a redundant sensor to sense the same process, which causes more frequent updates in the system we are considered, thus the AoI performance is better than the M/M/1/1 state update queue.
In the M/M/2 preemptive status updating queues, the newly collected updates in the system will preempt the oldest updates, which makes each sensor busy, while the system we consider is non-preempted FCFS, which leads to the possibility that the sensors will be idle, thus the average AoI of the system we consider is less than that of the M/M/2 preemptive status updating queues.

\section{Conclusion}
In this work, we studied the average AoI of an IoT-based remote monitoring system, where two sensors sample the same physical process and transmit status updates to a monitor.
We modeled the system as two M/M/1/1 parallel state update queues and derive general results for the average AoI based on the SHS framework.
We obtained closed-form expressions of average AoI for cases when the two sensors have the same the status arrival rates and/or the service rates.
We conducted simulations to verify our theoretical derivations.
It is confirmed that the average AoI of the two-sensor status update system is substantially reduced compared to that of the single sensor update system.

\bibliography{1570810246} 
\end{document}